# An ecologically valid examination of event-based and time-based prospective memory using immersive virtual reality: the effects of delay and task type on everyday prospective memory.

Panagiotis Kourtesis[a,b,c,d,e,f,g,h]*, Simona Collina[c,d], Leonidas A.A. Doumas[b], and Sarah E. MacPherson[a,b]

[a]Human Cognitive Neuroscience, Department of Psychology, University of Edinburgh, Edinburgh, UK;

[b]Department of Psychology, University of Edinburgh, Edinburgh, UK;

[c]Lab of Experimental Psychology, Suor Orsola Benincasa University of Naples, Naples, Italy;

[d]Interdepartmental Centre for Planning and Research "Scienza Nuova", Suor Orsola Benincasa University of Naples, Naples, Italy;

[e]National Research Institute of Computer Science and Automation, INRIA, Rennes, France;

[f]Univ Rennes, Rennes, France;

[g]Research Institute of Computer Science and Random Systems, IRISA, Rennes, France;

[h]French National Centre for Scientific Research, CNRS, Rennes, France.

* Panagiotis Kourtesis, Department of Psychology, University of Edinburgh, 7 George Square, Edinburgh, EH8 9JZ, United Kingdom. Email: pkourtes@exseed.ed.ac.uk

# Abstract

Recent research has focused on assessing either event- or time-based prospective memory (PM) using laboratory tasks. Yet, the findings pertaining to PM performance on laboratory tasks are often inconsistent with the findings on corresponding naturalistic experiments. Ecologically valid neuropsychological tasks resemble the complexity and cognitive demands of everyday tasks, offer an adequate level of experimental control, and allow a generalisation of the findings to everyday performance. The Virtual Reality Everyday Assessment Lab (VR-EAL), an immersive virtual reality neuropsychological battery with enhanced ecological validity, was implemented to comprehensively assess everyday PM (i.e., focal and non-focal event-based, and time-based). The effects of the length of delay between encoding and initiating the PM intention and the type of PM task on everyday PM performance were examined. The results revealed that everyday PM performance was affected by the length of delay rather than the type of PM task. The effect of the length of delay differentially affected performance on the focal, non-focal, and time-based tasks and was proportional to the PM cue focality (i.e., semantic relationship with the intended action). This study also highlighted methodological considerations such as the differentiation between functioning and ability, distinction of cue attributes, and the necessity of ecological validity.

## Introduction

Prospective memory (PM) involves the ability to remember to initiate an action in the future. The PM action may be associated with either a certain event (event-based e.g., when you see Alex, give him this card) or a specific time (time-based e.g., at 3 pm call Anne). The importance of PM tasks in everyday life varies from minimal importance (e.g., watering the plants every second day) to extreme importance (e.g., taking blood-pressure tablets after lunch). Hence, PM is paramount to various aspects of everyday life.

### *Event-based prospective memory*

The two predominant theoretical models of event-based PM are the preparatory attentional and memory (PAM) theory and the multiprocess theory (Anderson, McDaniel, & Einstein, 2017). The PAM theory suggests that efficient PM ability stems from a constant top-down monitoring of environmental and internal cues, which allow the individual to recall the intended action and initiate it (Smith, 2003; Smith, Hunt, McVay, & McConnell, 2007). The multiprocess theory is complementary to the PAM theory (McDaniel & Einstein, 2000; McDaniel & Einstein, 2007). In addition to PAM's top-down monitoring, the multiprocess theory postulates that effective performance on PM tasks is also facilitated by bottom-up spontaneous retrieval (McDaniel & Einstein, 2000; McDaniel & Einstein, 2007). Hence, when retrieving the intention to perform a PM task, the PAM theory suggests that there is an intentional retrieval process through the constant monitoring of target cues, while the multiprocess theory suggests an additional reflexive associative retrieval process by passively detecting environmental and internal cues (McDaniel, Umanath, Einstein, & Waldum, 2015). Bottom-up retrieval is thought to take place when the PM event-based target cue is focal, while top-down retrieval is implemented when the event-based target cue is non-focal (McDaniel & Einstein, 2000; McDaniel & Einstein, 2007; McDaniel *et al.*, 2015). More recently, the dynamic multiprocess framework has been proposed, which suggests a dynamic interplay between top-down and bottom-up retrieval processes for performing PM event-based tasks regardless of the focality of the target cue (Scullin, McDaniel, & Shelton, 2013; Shelton & Scullin, 2017).

*Focal event-based tasks*

A focal event-based PM task has a target cue with high salience (i.e., it can be easily detected) and high focality (i.e., it is strongly associated with the intended action; McDaniel & Einstein, 2000; McDaniel & Einstein, 2007; McDaniel *et al.*, 2015). For example, if an individual wants to buy a loaf of bread after work, while driving home, they may passively notice a large and easily detectable billboard at the side of the road (i.e., high salience), which displays pastry products (i.e., high focality). This focal target cue reminds the individual of their PM intention to buy a loaf.

*Non-focal event-based tasks*

In contrast, a non-focal event-based PM task has a target cue with low salience (i.e., it cannot be easily detected) and low focality (i.e., it is not strongly associated with the intended action; McDaniel & Einstein, 2000; McDaniel & Einstein, 2007; McDaniel *et al.*, 2015). Revisiting the above example, while driving home, the individual intentionally monitors the environment and detects a sign among other signs (i.e., low salience) for a supermarket (i.e., low focality). This non-focal cue detection assists the individual in retrieving the PM intention to buy a loaf.

*Performance differences*

Individuals have been found to perform substantially better on focal PM tasks than non-focal PM tasks. This indicates the difference in difficulty between the intentional and reflexive associative retrieval of the intention for each type of PM task (Anderson *et al.*, 2017; McDaniel et al., 2015; Mullet, Scullin, Hess, Scullin, Arnold, & Einstein, 2013; Scullin, McDaniel, Shelton, & Lee, 2010). Higher focality of the PM cue is thought to facilitate better encoding and retrieval of the PM intention (Anderson et al., 2017; McDaniel & Einstein, 2000; McDaniel & Einstein, 2007; McDaniel et al., 2015), which may explain the performance differences on focal and non-focal event-based tasks.

*Time-based prospective memory*

As stated above, everyday PM consists of both event- and time-based tasks (Einstein & McDaniel, 1996). Hence, including only event-based tasks in the assessment of PM does not allow for generalisation of the results to daily life (e.g., Einstein, Smith, McDaniel, & Shaw, 1997; Einstein, McDaniel, Manzi, Cochran, & Baker, 2000; Mullet et al., 2013; Scullin et al., 2010; Smith, 2003; Smith et al., 2007).

In the same way as the non-focal event-based tasks, the time-based tasks require individuals to monitor the environment for the appropriate target cue (Zuber, Mahy, & Kliegel, 2019). The target cue (e.g., a clock or a timer) is frequently non-salient (i.e., in the periphery or outside the field of view), while the focality of the cue is substantially low (i.e., the association between the time and intended PM action). Time-monitoring is also important for performing time-based tasks (Glickson, & Myslobodsky, 2006). Individuals appear to implement strategic time monitoring where, by checking the time, they estimate when they should repeat their time check and correspondingly regulate the frequency of their time monitoring (Cona, Arcara, Tarantino, & Bisiacchi, 2012; Kliegel, Martin, McDaniel, & Einstein, 2001; Mäntylä, Carelli, & Forman, 2007). For example, if an individual is required to take a blood-pressure tablet at 8 pm, they will monitor the time on their watch, which reminds them of the intended PM action, and will intensify their time monitoring as the appropriate time approaches. When the time is approximately 8 pm, the individual will perform the PM intended action and take the blood-pressure tablet.

The literature suggests that time-based tasks are more cognitively demanding than event-based tasks as individuals frequently perform more poorly on time-based tasks compared to event-based tasks (Einstein & McDaniel, 1996). In laboratory tasks, often performance on both focal and non-focal tasks is significantly better than performance on time-based tasks (e.g., Conte & McBride, 2018; Zuber et al., 2019). However, in some lab-based studies, only focal task performance was found to be significantly better than performance on time-based tasks (e.g., McBride & Flaherty, 2020).

*The importance of ecological validity*

Many of the studies examining performance on focal, non-focal, and time-based PM tasks have involved lab-based paradigms (e.g., Conte & McBride, 2018; Einstein et al., 1997, 2000; McBride & Flaherty, 2020; Mullet et al., 2013; Scullin et al., 2010; Smith, 2003; Smith et al., 2007; Zuber et al., 2019). These laboratory tasks incorporate simple, static stimuli within a highly controlled environment and do not resemble the complexity of real-life situations (Parsons, 2015). For example, a laboratory PM experiment may incorporate a single on-going task such as a computerised lexical decision task, where the participant should indicate whether the presented word is an actual or fictional word, with a comparable computerised PM task, where the participant should indicate when a specific word appears. Hence, the laboratory tasks incorporate simple stimuli (e.g., static) and testing environments (e.g., two-dimensional), almost automatic responses (i.e., in milliseconds) provided using a keyboard, a relatively short testing duration (e.g., 15 minutes) and delay after encoding a PM intention (e.g., up to 10 minutes), and a single on-going task and a low difficulty PM task.

PM studies adopting experimental tasks and conditions that substantially diverge from the complexity and cognitive demands of everyday tasks may provide discrepant results compared to ecologically valid research paradigms (Chaytor & Schmitter-Edgecombe, 2003; Franzen & Wilhelm, 1996; Marsh, Hicks, & Landau, 1998; Spooner & Pachana, 2006). Implementing ecologically valid tasks allows one to generalise the findings to everyday functioning (Chaytor & Schmitter-Edgecombe, 2003; Haines *et al.*, 2019; Higginson, Arnett, & Voss, 2000; Mlinac & Feng, 2016; Parsons, 2015; Phillips, Henry, & Martin, 2008). Therefore, adopting ecologically valid research paradigms allows researchers to understand PM performance in daily life.

*Naturalistic Experiments*

Ecologically valid research paradigms have predominantly involved assessments in various real-life settings such as performing errands in a shopping centre or a pedestrianized street or at home (e.g., Garden, Phillips, & MacPherson, 2001; Marsh *et al.*, 1998; Shallice & Burgess, 1991). In PM research, a discrepancy is found between performance on laboratory tasks and naturalistic tasks in terms of the so-called age-prospective memory-paradox (Kvavilashvili, Cockburn, & Kornbrot, 2013; Niedźwieńska & Barzykowski, 2012; Phillips *et al.,* 2008; Uttl, 2008). In laboratory tasks, there are significant age-related differences found on event-based and time-based tasks where older adults perform more poorly than younger adults, while in naturalistic tasks, age differences are not typically found (Kvavilashvili *et al.*, 2013; Niedźwieńska & Barzykowski, 2012; Phillips *et al.,* 2008; Uttl, 2008). In naturalistic tasks, older adults may use compensatory strategies and external memory aids to successfully perform PM tasks (Phillips *et al.,* 2008; Shelton & Scullin, 2017; Uttl, 2008). However, naturalistic experiments rarely consider the diverse types of PM tasks within the same paradigm. Niedźwieńska and Barzykowski (2012) investigated performance on focal, non-focal, and time-based tasks in a naturalistic setting and found no significant differences among the PM tasks. In contrast, the same comparisons using lab-based tasks revealed that, in younger adults, performance on the event-based task with focal cue was performed better than the event-based task with non-focal cue and the time-based PM task, whereas performance on these latter two tasks did not differ. In older adults, performance on the event-based task with non-focal cue was significantly poorer than on both the event-based task with focal cue and the time-based task, while performance on these latter two tasks did not differ.

Nonetheless, real-world tasks suffer from several limitations. These include a lack of standardization for use in other clinics and laboratories, reduced feasibility for use with certain neurological populations (e.g., schizophrenia patients), being time-consuming and expensive, having procedural and bureaucratic demands, as well as diminished experimental control over external factors (Logie, Trawley, & Law, 2011; Parsons, 2015).

*Virtual Reality*

Another approach for achieving ecological validity involves the implementation of technological mediums such as recordings of real-world locations and non-immersive virtual environments (Farrimond, Knight, & Titov, 2006; Logie *et al.*, 2011; McGeorge *et al.*, 2001; Paraskevaides *et al.*, 2010). These technologies are able to simulate real-life tasks, are cost-effective, require less administration time, enable increased experimental control, and can be implemented in other clinical and research settings (Parsons, McMahan, & Kane, 2018; Werner & Korczyn, 2012; Zygouris & Tsolaki, 2015). Non-immersive virtual reality (VR) refers to the software (i.e., virtual environments) which are projected onto two-dimensional displays (i.e., hardware) such as computer screens, laptops, tablets, and mobile devices (Kourtesis, Collina, Doumas, & MacPherson, 2019a). Non-immersive VR tests, such as the VR Prospective Memory Test (Man et al., 2018) and the Virtual Ride in a Town (Lecouvey et al., 2019), have been implemented in an attempt to assess PM. However, non-immersive VR tests can be challenging for individuals without gaming backgrounds (Zaidi, Duthie, Carr, & Maksoud, 2018), especially for older adults and clinical populations (e.g., individuals with mild cognitive impairment or Alzheimer's disease; Werner & Korczyn, 2012; Zygouris & Tsolaki, 2015).

In contrast, immersive VR refers to virtual environments (i.e., software) which are projected onto devices such as head-mounted displays (HMDs) and CAVE systems (Kourtesis et al., 2019a). Non-immersive VR provide a less ecologically valid testing environment than immersive VR tests (Bohil, Alicea, & Biocca, 2011; Parsons, 2015; Rizzo, Schultheis, Kerns, & Mateer, 2004; Teo *et al.*, 2016). Importantly, differences in performance between gamers and non-gamers are mitigated by the first-person view and ergonomic interactions that are proximal to real-life actions in immersive VR environments (Zaidi *et al.*, 2018). Consequently, an immersive VR research paradigm provides the most efficient approach to study everyday PM performance.

**The Virtual Reality Everyday Assessment Lab**

We have recently developed the Virtual Reality Everyday Assessment Lab (VR-EAL), which is an immersive VR neuropsychological battery assessing everyday cognitive functions such as PM (i.e., event-based and time-based), episodic memory (i.e., immediate and delayed recognition), attentional processes (i.e., visual, visuospatial, and auditory), and executive functioning (i.e., planning and task-shifting; Kourtesis, Korre,

Collina, Doumas, & MacPherson, 2020b). The convergent, construct, and ecological validity of the VR-EAL has been provided against established ecologically valid paper-and-pencil tests such as the Cambridge Prospective Memory Test (Wilson *et al.*, 2005), the Rivermead Behavioral Memory Test–III (Wilson, Cockburn, & Baddeley, 2008), the Test of Everyday Attention (Robertson, Ward, Ridgeway, & Nimmo-Smith, 1996), and the Behavioral Assessment of Dysexecutive Syndrome (Wilson, Alderman, Burgess, Emslie, & Evans, 1997; Kourtesis, Collina, Doumas, & MacPherson, 2020a). Notably, the VR-EAL does not induce differences in performance between gamers and non-gamers, and the VR-EAL tasks were rated by participants as substantially more similar to everyday tasks compared to their equivalent ecologically valid paper-and-pencil tasks (Kourtesis *et al.*, 2020a). Hence, the VR-EAL better resembles the complexity and cognitive demands of real-life tasks.

The VR-EAL assesses both event-based and time-based PM (Kourtesis *et al.*, 2020b). It incorporates five event-based tasks and four time-based tasks. The event-based tasks consist of two non-focal event-based tasks, two focal event-based tasks, and one misleading task (i.e., prompting the examinee to perform a task when there is no task to perform). In addition, the focal and non-focal event-based tasks are further subdivided into tasks performed at an early or late stage of the scenario. Similarly, the time-based tasks are divided into early (two tasks) and late tasks (two tasks), where one of the two late tasks is a misleading task. The segregation of the PM tasks into early and late tasks allows us to investigate the effect of the length of the delay between encoding the PM intention and performing the PM action.

Moreover, PM assessment in the VR-EAL is aligned with the methodological suggestions for the assessment of event-based PM provided by McDaniel and collaborators (2015). There are only two focal event-based tasks within a scenario of approximately 70 minutes. The focal event-based tasks have a single focal cue each and there are no other cues driving the participant's attention to the PM task. Participants are instructed that the PM tasks need to be performed later in the scenario (i.e., anytime between 15 and 60 minutes after forming a PM intention). Finally, the importance of PM and non-PM tasks (e.g., planning the itinerary, preparing breakfast, and buying groceries from the supermarket) is equally highlighted. The non-PM VR-EAL tasks do not serve simply as distractive on-going tasks but are cognitively demanding tasks that assess cognitive functions central to everyday functioning.

### *The current study*

To our knowledge, the present study is the first to adopt an ecologically valid paradigm using immersive VR to examine real-life PM. In addition, our study is the first to assess, in a single sample, everyday PM performance on focal event-based, non-focal event-based and time-based PM tasks with diverse delay lengths using an ecologically valid immersive VR paradigm. Previous studies have found a significant effect of delay on either focal event-based tasks (e.g., Kliegel & Jager, 2006; Meier, Zimmermann, & Perrig, 2006; Scullin & McDaniel, 2010), non-focal event-based tasks (e.g., Martin, Brown, & Hicks, 2011; McBride, Beckner, & Abney, 2011), or time-based tasks (e.g.,

McBride, Coane, Drwal, & LaRose, 2013). However, these significant performance differences between the various types of PM tasks in laboratory experiments have not been replicated using naturalistic paradigms, except in McBride et al. (2013) which involved time-based tasks.

In this study, we have implemented the VR-EAL, which has been validated and evaluated as being more ecologically valid against standardized ecological PM measures. We also subdivided everyday PM performance into PM functioning and PM ability. PM functioning is based on the VR-EAL PM score, which provides values from 0 – 6 proportionally to the number of prompts (i.e., external aids) that the participant required to successfully perform the PM task. PM ability is based on a dichotomised VR-EAL PM score (i.e., 0 or 1) where zero indicates a failure to perform the PM task without using external aids, and 1 indicates success in performing the PM task without using external aids.

Here, we explored the effects of delay and PM task type on everyday PM functioning and ability. Since an effect of PM task type has not been observed in naturalistic settings (e.g., Niedźwieńska & Barzykowski, 2012), we expected that the effect of delay on PM performance would be more important than the effect of task type on PM performance. However, previous research has shown that the effect of delay is greater on non-focal event-based (Martin *et al.*, 2011; McBride *et al.*, 2011) and time-based tasks (McBride *et al.*, 2013), which may be attributed to the focality of the target cue (Martin *et al.*, 2011; McBride *et al.*, 2011, 2013). Since the focality of the target cue in each PM task type varies, we expected an effect of delay on the performance on each PM task type. We expected that the effect of delay would be greater on time-based tasks compared to the event-based tasks. Equally, we expected that the effect of delay would be greater on non-focal event-based tasks than focal event-based tasks.

Furthermore, we investigated the performance on each type of PM task, as well as on PM tasks with a diverse length of delay. Considering the enhanced ecological validity of the VR-EAL and the discrepancies that have emerged from naturalistic experiments, we hypothesized that there would not be significant differences between performance on the PM tasks, unlike the differences found using laboratory-based studies. On the other hand, we expected significant differences between early and late PM tasks. Specifically, everyday PM performance on substantially delayed PM tasks was anticipated to be significantly decreased, especially for time-based and non-focal event-based tasks.

**Methods**

*Participants and Procedure*

Participants were recruited via social media and the internal mailing list of the University of Edinburgh and were the same cohort as in Kourtesis *et al.* (2020a). Forty-one participants (21 females) with a mean age of 29.15 years (SD = 5.80, range = 18-45) and mean education of 13.80 years (SD = 2.36, range = 10-16) were recruited. The study was approved by the Philosophy, Psychology and Language Sciences Research Ethics Committee of the University of Edinburgh. Written informed consent was obtained from each participant. All participants received verbal and written instructions regarding the procedures, possible adverse effects of immersive VR (e.g., cybersickness), utilization of the data, and general aims of the study. The participants attended the session individually. Upon their arrival, they were instructed by the examiner how to use the VR equipment. Then the participants were immersed into VR-EAL, where they had a practice period without any time restrictions to familiarise themselves with the basic controls and configuration. When the participants felt comfortable interacting with the virtual environment, the VR-EAL scenario commenced. All participants completed all tasks (i.e., non-PM and PM) and the full-VR-EAL scenario. See below a description of the VR-EAL's tasks and scenario. Finally, after exposure to VR, the Virtual Reality Neuroscience Questionnaire (VRNQ; Kourtesis, Collina, Doumas, &MacPherson, 2019b) was used to assess cybersickness symptomatology.

*Materials*

*Hardware*

An HTC Vive HMD with two lighthouse stations for motion tracking and two HTC Vive wands with six degrees of freedom (6DoF) for navigation and interactions within the virtual environment were implemented in accordance with our previously published technological recommendations for immersive VR research (Kourtesis et al., 2019a). The spatialized (bi-aural) audio was facilitated by a pair of Senhai Kotion Each G9000 headphones. The size of the VR area was $5m^2$, which provides an adequate space for immersion and naturalistic interaction within virtual environments (Borrego, Latorre, Alcañiz, & Llorens, 2018). The HMD was connected to a laptop with an Intel Core i7 7700HQ 2.80GHz processor, 16 GB RAM, a 4095MB NVIDIA GeForce GTX 1070 graphics card, a 931 GB TOSHIBA MQ01ABD100 (SATA) hard disk, and Realtek High Definition Audio.

*VR-EAL*

VR-EAL can be run on any VR HMD which is compatible with SteamVR, such as HTC Vive series (e.g., Pro, Pro Eye, and Cosmos), Oculus Rift series (e.g., Rift and Rift S), Pimax series (e.g., 4K, 5K, and 5K Plus), Varjo series (e.g., VR-1 and VR-2), Samsung Odyssey series (e.g., Odyssey and Odyssey +), and Valve Index. VR-EAL assesses everyday cognitive functions such as PM, episodic memory (i.e., immediate and delayed recognition), executive functioning (i.e., planning, multitasking) and selective visual, visuospatial and auditory (bi-aural) attention within a realistic immersive VR scenario lasting around 70 minutes (Kourtesis *et al.*, 2020a). See Table 1 and Figures 1 and 2 for a summary of the VR-EAL tasks assessing each cognitive ability. In addition, a brief video recording of the VR-EAL may be accessed at this hyperlink:

https://www.youtube.com/watch?v=IHEIvS37Xy8&t . Please see supplementary material for a more comprehensive description of the VR-EAL tasks that are not considered in the current study (i.e., ongoing tasks, and misleading PM tasks).

Table 1. VR-EAL tasks and score ranges

| Scene | Cognitive Function | Task | Score |
|---|---|---|---|
| 3 | Prospective memory | Write down the notes for the errands. | 0 – 6 |
| 3 | Immediate recognition | Recognising items on the shopping list. | 0 – 20 |
| 3 | Planning | Drawing the route to be taken. | 0 – 19 |
| 6 | Multitasking | Cooking task (preparing breakfast). | 0 – 16 |
| 6 | Prospective memory – event based | Take medication after breakfast. | 0 – 6 |
| 8 | Selective visuospatial attention | Collect items from the living room. | 0 – 20 |
| 8 | Prospective memory – event based | Take the chocolate pie out of the oven. | 0 – 6 |
| 10 | Prospective memory – time based | Call Rose at 10 am. | 0 – 6 |
| 12 | Selective visual attention | Find posters on both sides of the road. | 0 – 16 |
| 14 | Delayed recognition | Recognising items from the shopping list. | 0 – 20 |
| 15 | Prospective memory – time based | Collect the carrot cake from the bakery at 12 pm. | 0 – 6 |
| 16 | Prospective memory – event based | False prompt before going to the library. | -6 – 0 |
| 17 | Prospective memory – event based | Return the red book to the library. | 0 – 6 |
| 19 | Selective auditory attention | Detect sounds from both sides of the road. | 0 – 32 |
| 20 | Prospective memory – time based | False prompt before going back home. | -6 – 0 |
| 21 | Prospective memory – event based | Back home, give the extra pair of keys to Alex. | 0 – 6 |
| 22 | Prospective memory – time based | Take the medication at 1pm. | 0 – 6 |

*The tasks are presented in the same order as they are performed within the scenario. The table derives from Kourtesis et al., (2020a).

Figure 1. VR-EA L Storyline: Scenes 3 - 12.

**Scene 3**      **Scene 3**

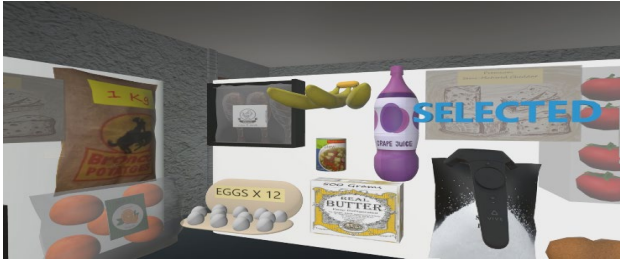 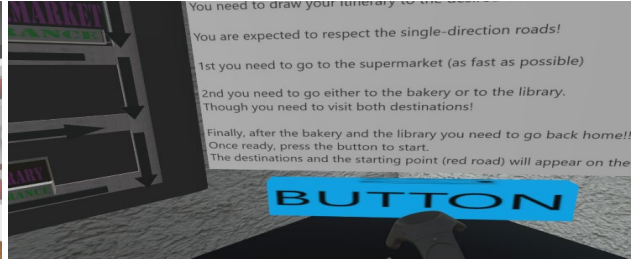

**Scene 6**      **Scene 6**

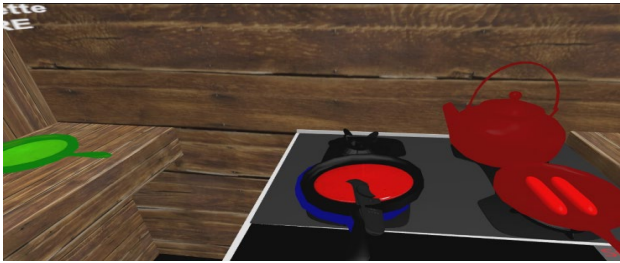 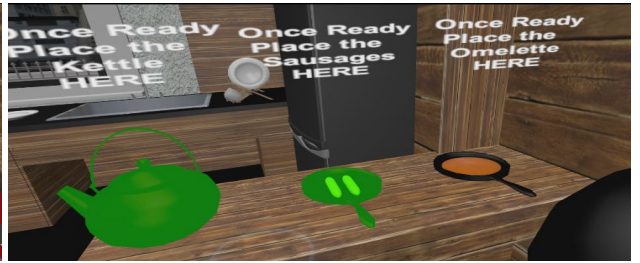

**Scene 6**      **Scene 8**

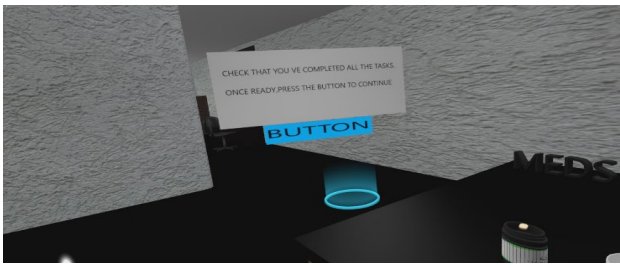 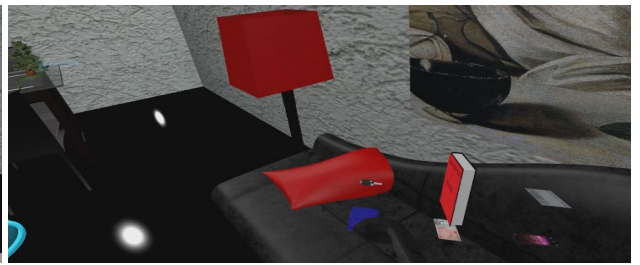

**Scene 8**      **Scene 10**

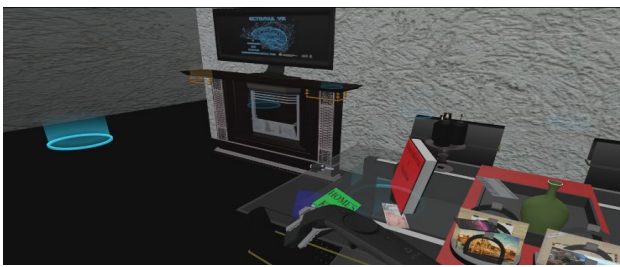 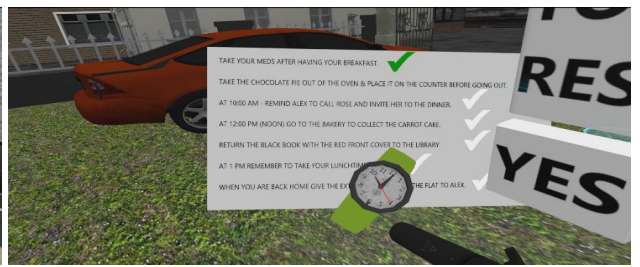

**Scene 12**      **Scene 12**

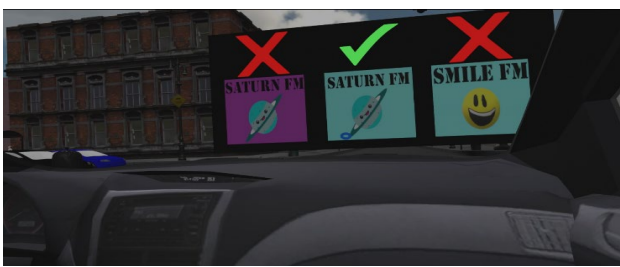 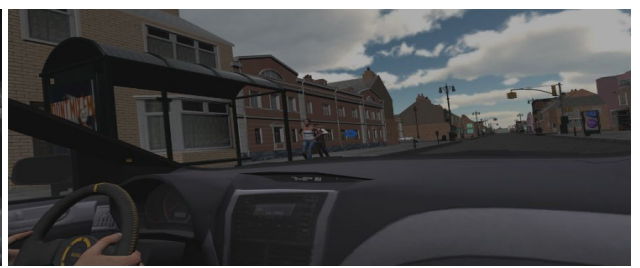

*Derived from Kourtesis et al., (2020b).*

Figure 2. VR-EAL Storyline: Scenes 14 – 22

**Scene 14**            **Scene 14**
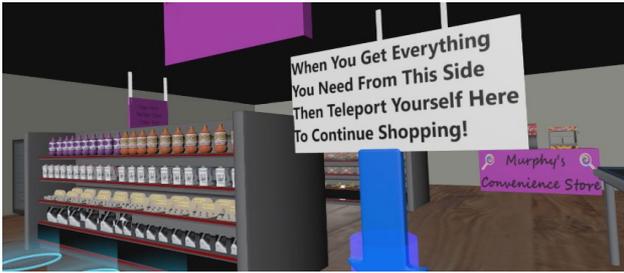    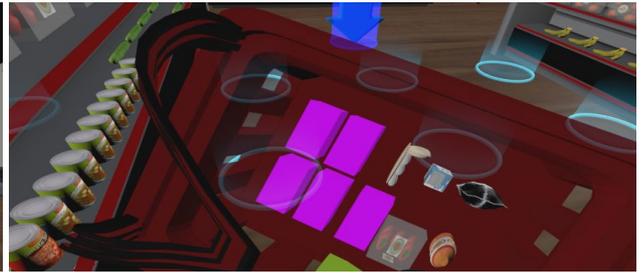

**Scene 15**            **Scene 17**
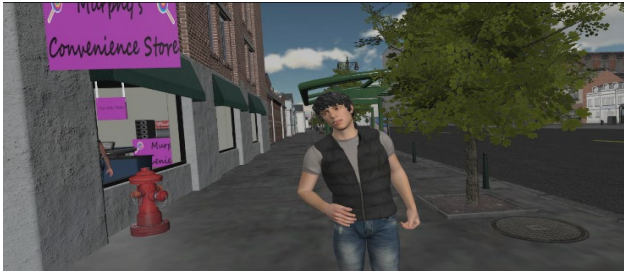    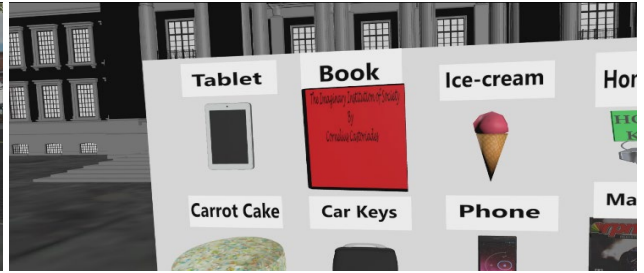

**Scene 19**            **Scene 19**
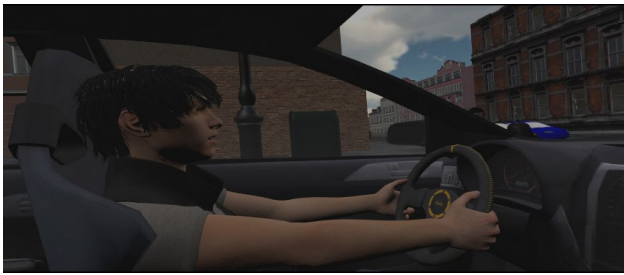    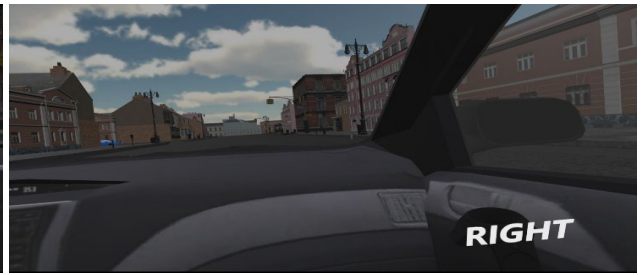

**Scene 20**            **Scene 22**
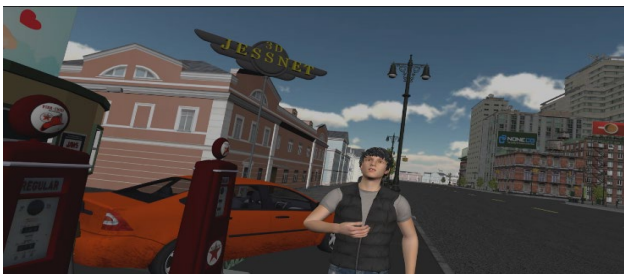    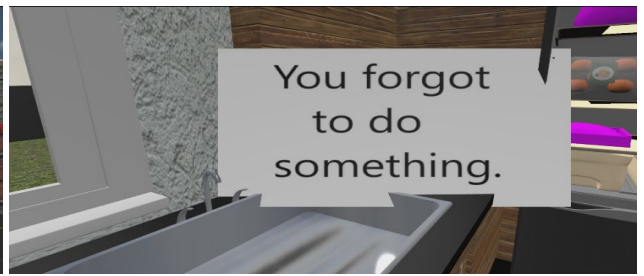

**Scene 22**            **Scene 22**
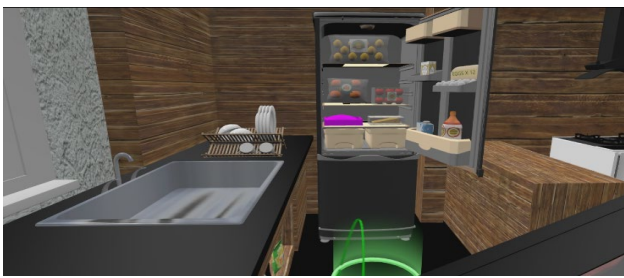    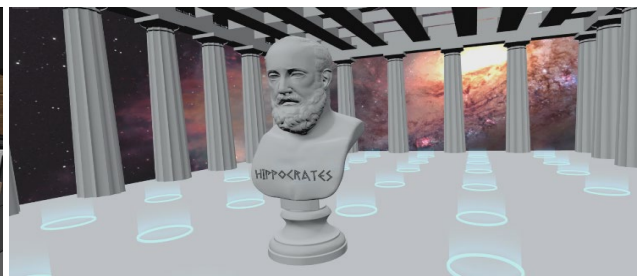

*Derived from Kourtesis et al., (2020b).*

*Prospective Memory.* VR-EAL considers both event-based and time-based PM tasks. The PM tasks are divided into five event-based tasks and four time-based tasks. In the event-based tasks, the participant should remember to perform a PM action when a specific event occurs (e.g., take medicines after breakfast). In the time-based tasks, the examinee should remember to perform a planned action at a specific time (e.g., call Rose at 12 pm). Consider the following as a scoring example. At the end of a scene, the examinees should press a button to confirm that all the tasks in the scene are completed. If the examinees have already taken their medication (i.e., PM task) before pressing the final button, then the scene ends, and the examinees receive 6 points. Otherwise, the first prompt appears (i.e., "You Have to Do Something Else"). If the examinees then follow the prompt and take their medication, they receive 4 points. If the examinees press the final button again, then the second prompt appears (i.e., "You Have to Do Something After Having your Breakfast"). If the examinees follow this prompt and take their medication, they receive 2 points. If the examinees press the final button again, then the third prompt appears (i.e., "You Have to Take Your Meds"). If the examinees follow this prompt and take their medication, they then receive 1 point. If the examinees repress the final button without ever taking their medication, they get zero points, and the scene ends. Scores range from zero to six.

*Focal event-based tasks.* There are two focal event-based tasks, where the first task is performed in an early stage of the scenario (15-30 minutes after encoding the intention), and the second one is performed in a late stage (45-60 minutes after encoding the intention). The environmental cue in focal event-based tasks should be easily detected (i.e., salient) and of high focality (e.g., to be in the same semantic category). For example, the examinee is in front of the library facing the large sign displaying "Library", where the examinee should remember to return a book.
*Non-focal event-based tasks.* Comparable to the focal event-based tasks, there are two non-focal event-based tasks, which are performed in an early and late stage. However, in the non-focal event-based tasks, the focality and salience of the cue should be adequately low to motivate the examinee to monitor the area for cues. For example, the examinee finishes preparing breakfast (see the cooking task below), and after having breakfast, the examinee should remember to take his/her medicine (see the scoring example above). The medicine is out of the field of view of the examinee, but it is in the same room (i.e., kitchen) among other kitchen objects (e.g., cups, plates, towels, microwave, and toaster). Hence, the examinee should purposefully monitor the area to detect the medicine and perform the PM task.

*Time-based tasks.* There are four time-based tasks, from which three are traditional time-based tasks (e.g., call Rose at 12pm) and one is a misleading task (see below). The examinee should be checking his/her digital watch, which is attached to the left hand/controller in the virtual environment, to perform the time-based task. Hence, the salience of the PM cue (i.e., the time on the watch) is low, while the focality of the PM cue (i.e., association between time and the intended action) is extreme low as frequently happens in daily time-based tasks. Two of the traditional time-based tasks, one performed in an early stage (i.e., 15-30 minutes after encoding the intention) and one in a late stage (i.e., 45-60 minutes after encoding the intention), are used in the comparison with the focal and non-focal event-based tasks.

*Design and Variables*

This study follows a within subjects repeated measures design. The participants performed all the tasks while following the VR-EAL scenario as described above. There are focal, non-focal, and time-based tasks with different delay lengths between encoding and retrieval (i.e., performed at an early or a late stage of the scenario). Controlling the type of PM task and the length of delay allows for scrutiny of their effect on everyday PM performance. However, everyday PM performance is examined in terms of everyday PM functioning (i.e., considering the correct responses after receiving a prompt to perform that particular task) and everyday PM ability (i.e., the correct responses without requiring a prompt to perform that particular task).

*Everyday prospective memory performance.*

*Prospective memory functioning.* The VR-EAL PM score on each task is used for depicting everyday PM functioning. As described above, the VR-EAL's scoring system provides a PM score for each task ranging from zero to six. The score is proportional to the number of prompts that were required to perform the corresponding task, as well as the correctness of the PM response (i.e., performing the PM task that corresponds to this event or time).

*Prospective memory ability.* A dichotomous PM score is used for depicting everyday PM ability. Consequently, the possible values for this score are either 0 or 1. A dichotomous PM score equal to 1 corresponds to the VR-EAL PM score equal to 6, while zero corresponds to the remaining possible values for the VR-EAL PM score (i.e., 0 - 5). Hence, a dichotomous PM score equal to 1 signifies that the participant performed the respective PM task independently without requiring a prompt, while the 0 indicates that the participant required a prompt to perform the corresponding PM task.

*Prospective memory tasks' parameters*

*Type of prospective memory task.* This variable will be referred as PM task type, and it has three levels (i.e., focal, non-focal, and time-based). Two VR-EAL PM tasks of each type were considered. Thus, there are two focal, two non-focal, and two time-based tasks. The tasks were selected to have similar demands (i.e., difficulty) and in terms of their length of delay (see below). The similar demands across items was ensured during the development of VR-EAL. One pilot study was conducted for each of the three versions of VR-EAL (i.e., alpha, beta, and final) involving professional and non-professional user testers (Kourtesis et al., 2020b). During these studies, participants provided quantitative and qualitative feedback regarding the tasks and the scenario. The criteria included the proximity to the target (e.g., similar distances between the cue and performing the PM task), ergonomics (e.g., similar physical movements to perform the PM task), user interface (e.g., performing the PM tasks using button presses, while touching the relevant object), similar types of feedback (e.g., visual, auditory, and haptic), and time of performance (e.g., tasks were performed after interacting with the environment and/or 3D character). More information regarding the development of VR-EAL is provided in the Kourtesis and collaborators (2020b) study.

*Length of delay between encoding and retrieval.* This variable has two levels (i.e., early and late), where the early tasks are performed approximately 15-30 minutes after forming the PM intention (i.e., encoding), and the late tasks are performed approximately 45-60 minutes after forming the PM intention. One PM task of each type (i.e., focal, non-focal, and time-based) is performed at an early stage in the scenario, and the second task of each type is performed at a late stage in the scenario. This temporal differentiation allows the inspection of the delay's effect on PM performance for each type of PM task. This also facilitates a comparison between the performance on early and late PM tasks per type (i.e., focal event-based, non-focal event-based, and time-based tasks).

### Statistical Analyses

An *a priori* power analysis for an $\alpha = .05$ and a power $\geq 80\%$ indicated that a sample size of $N \geq 32$ is required to detect moderate effects in regression analyses and post hoc tests. Our sample size (N=41) is well beyond this threshold. All the statistical analyses were conducted using R language (R Core Team, 2020) in RStudio (RStudio Team, 2020). The distribution of the performance variables was initially examined. The distribution of the VR-EAL PM scores (i.e., everyday PM functioning) followed a Poisson distribution, while the distribution of the dichotomous PM score (i.e., everyday PM ability) followed a binomial distribution.

*Mixed effects generalised linear regression analyses*

Mixed effects generalised linear regression analyses were performed to investigate which parameters affect and predict everyday PM functioning and ability. The mixed effects generalised linear regression analyses were performed using the stats (RStudio Team, 2020) and lmerTest packages (Kuznetsova, Brockhoff, and Christensen, 2017). Corresponding to the distribution family of each performance variable, mixed effects generalised linear regression models were created. Firstly, a model for each fixed effect (i.e., PM task type and Delay) and the mixed effect (i.e., the interaction between PM task type and Delay) were created. The random effect per participant was also included in each model. A null model was also created including only the random effect per participant. The models were compared by conducting analyses of deviance. The criteria for the comparisons between the models were the Akaike's information criterion (AIC), the deviance of the model, and the $\chi^2$ probability. Initially, the three models (i.e.,

PM task type, Delay, and mixed effects) were compared individually with the null model. Then, the 3 models were compared with one another. The best model among them was used for building the models in the next step by adding predictors to it. An incremental approach was adopted, where a variable was added as a predictor to the current best model. Each new model was then compared against the current best model and the null model. After identifying the model that has the best combination of predictors, this model was then used for adding variables as random effects per participant. As described above, the same incremental approach with stepwise comparisons against the null model and the current best model was implemented for the random effects. The model that was identified as the best model was then accepted as the model that best predicts PM functioning or ability.

*Post hoc comparisons for the models.* To further explore the effect of each parameter on everyday PM functioning and ability, the least-squares means of the models' predictors were calculated and then compared. The post hoc comparisons of the predictors' least-squares means were performed using the emmeans package (Lenth, 2020). The p-values of the comparisons were adjusted using a Bonferroni correction to avoid p-value inflation due to the existence of multiple comparisons.

*Pairwise multiple comparisons of mean rank sums*

Non-parametric multiple comparisons based on the mean ranks sums of the observations were performed to examine the performance differences in PM functioning and ability under each condition pertinent to PM task type, the length of delay, and the possible combinations between them. The Durbin-Conover test and the Wilcoxon mean rank sum test were implemented for the variables with 3 (i.e., PM task type) and 2 (i.e., length of delay) levels respectively. The p-values were corrected for possible inflations due to the multiple comparisons using the Bonferroni correction method. The PMCMR (Pohlert, 2014) and the ggstatsplot (Patil, 2018) packages were used for conducting and visualising the pairwise multiple comparisons of mean rank sums.

**Results**

The spectrum of VRNQ possible responses on cybersickness items include 1-extremely intense feeling, 2-very intense feeling, 3-intense feeling, 4-moderate feeling, 5-mild feeling, 6-very mild feeling, and 7-absent feeling. In the current study, the medians for nausea, dizziness, disorientation, and instability items of VRNQ were 7 (i.e., absent feeling), except for fatigue, which was 6 (i.e., very mild feeling). Only intense cybersickness symptoms appear to affect cognitive performance, while moderate symptoms do not appear to affect it (Mittelstaedt, Wacker, & Stelling, 2019). Notably, no participant in our study reported cybersickness symptoms (i.e., nausea, dizziness, disorientation, fatigue, or instability) stronger than a mild feeling, which postulates that cybersickness symptoms did not affect the participants' performance.

The descriptive statistics for the PM scores are displayed in Table 2. In terms of everyday PM functioning, the length of the delay was found to better predict performance than the type of PM task. Nevertheless, together these factors only explained 5% of the observations and random effects were found to significantly affect everyday PM functioning. In contrast, in terms of everyday PM ability, random effects did not significantly affect performance. The length of delay again explained everyday PM performance better than the PM task type. The interaction between the length of delay and the PM task type was found to better explain everyday PM ability, where the interaction explained 27% of the observations pertinent to everyday PM ability. The post-hoc comparisons of the least-squares means and the mean ranks sums indicated significant differences between the tasks performed at an early (i.e., 15 – 30 mins) and late (i.e., 45 – 60 mins) delay within the scenario for both everyday PM functioning and ability. There were no significant performance differences between the types of PM tasks when the length of delay was 15 – 30 minutes. However, significant differences between the types of PM tasks were found when the length of delay was more than 45 minutes.

Table 2. Descriptive statistics for the PM scores

| PM Task Type | Score | Delay | Mean (SD) | Median | Range |
|---|---|---|---|---|---|
| Focal | VR-EAL (max = 6) | All | 5.34 (1.00) | 6.0 | 2 - 6 |
| | Dichotomous (max = 1) | All | 1.00 (0.00) | 1.0 | 0 - 1 |
| Non-Focal | VR-EAL (max = 6) | All | 4.95 (1.01) | 5.0 | 2 - 6 |
| | Dichotomous (max = 1) | All | 0.50 (0.50) | 0.5 | 0 - 1 |
| Time-based | VR-EAL (max = 6) | All | 4.57 (1.31) | 4.0 | 1 - 6 |
| | Dichotomous (max = 1) | All | 0.39 (0.49) | 0.0 | 0 - 1 |
| All | VR-EAL (max = 6) | Early | 5.29 (1.11) | 6.0 | 1 - 6 |
| | Dichotomous (max = 1) | Early | 0.68 (0.47) | 1.0 | 0 - 1 |
| All | VR-EAL (max = 6) | Late | 4.62 (1.15) | 4.0 | 2 - 6 |
| | Dichotomous (max = 1) | Late | 0.37 (0.48) | 0.0 | 0 - 1 |
| Focal | VR-EAL (max = 6) | Early | 5.46 (1.00) | 6.0 | 2 - 6 |
| | | Late | 5.22 (0.99) | 6.0 | 4 - 6 |
| | Dichotomous (max = 1) | Early | 0.76 (0.44) | 1.0 | 0 - 1 |
| | | Late | 0.61 (0.49) | 1.0 | 0 - 1 |
| Non-Focal | VR-EAL (max = 6) | Early | 5.27 (1.07) | 6.0 | 2 - 6 |
| | | Late | 4.63 (1.04) | 4.0 | 2 - 6 |
| | Dichotomous (max = 1) | Early | 0.66 (0.48) | 1.0 | 0 - 1 |
| | | Late | 0.34 (0.48) | 0.0 | 0 - 1 |
| Time-based | VR-EAL (max = 6) | Early | 5.15 (1.26) | 6.0 | 1 - 6 |
| | | Late | 4.00 (1.10) | 4.0 | 2 - 6 |
| | Dichotomous (max = 1) | Early | 0.63 (0.49) | 1.0 | 0 - 1 |
| | | Late | 0.15 (0.36) | 0.0 | 0 - 1 |

N = 41; "Early" and "Late" corresponds to the time in the scenario that this PM task was performed. Dichotomous corresponds to dichotomized outcomes i.e., 0 = failure to recall PM intention and 1 = successfully recalled PM intention.

*Mixed effects generalised linear regression models*

The models with additional variables as random effects did not appear better than the equivalent models with only a random effect per participant, while the latter models had better AIC and deviance.

*Everyday prospective memory functioning*

The models for everyday PM functioning are displayed in Table 3. Compared to the null model, the models containing only the length of delay, the PM task type, or the interaction between them as predictors were found to explain everyday PM functioning significantly better. The delay model appeared to have the better fit to the data based on AIC, while the interaction model explained a greater percentage of the observations. However, the best model that emerged was the model with both PM task type and the length of delay as predictors of everyday PM functioning, indicating that the PM task type and the length of delay affect PM functioning. Nevertheless, the model explained only 5% of everyday PM functioning.

Table 3. Mixed effects generalised linear regression models for VR-EAL's PM score.

| Model | F | $\chi^2$ | p-value ($> \chi^2$) | AIC | Deviance | $R^2$ |
|---|---|---|---|---|---|---|
| Null (only random effects) | - | - | - | 925.15 | 921.15 | 0.00 |
| Delay | 5.64 | 05.66 | .017* | 921.49 | 915.49 | 0.02 |
| PM task type | 2.44 | 4.89 | .087 | 924.26 | 916.26 | 0.02 |
| Interaction between Delay and PM task type | 2.43 | 12.69 | .026* | 922.46 | 908.46 | 0.06 |
| *Delay and PM task type* | *5.64 2.44* | *10.54* | *.014** | *920.61* | *910.61* | *0.05* |

The best model is displayed in *Italics*; AIC = Akaike's Information Criterion; For AIC and Deviance, a smaller value indicates a better fit for the model; The displayed $\chi^2$ statistics are against the null model; $R^2$ is a pseudo-$R^2$ for mixed effects generalised linear regression models.

The effects on everyday PM functioning are presented in Table 4. The random effect per participant was found to be significant, postulating the existence of individual differences affecting everyday PM functioning. In addition, a greater length of delay appears to significantly negatively affect PM functioning, while the inverse is observed for shorter delays. Similarly, the focal event-based tasks appeared to increase PM functioning, while the time-based tasks were found to decrease everyday PM functioning. This postulates that the focality and the salience of the cue indeed affect everyday PM functioning. Finally, the interaction between PM task type and length of delay appears to have an effect on PM functioning, except for the late non-focal combination.

Table 4. Effects on VR-EAL's PM Score.

| Effect | Levels | z Value | p-value (>\|z\|) | β coefficient | β 95% CI |
|---|---|---|---|---|---|
| *Random effect per participant* | - | *55.88* | *<.001\*\*\** | *1.60* | *[1.54, 1.66]* |
| *fixed effect per Delay* | *Early* | *42.52* | *<.001\*\*\** | *1.67* | *[1.59, 1.74]* |
| | *Late* | *-2.38* | *.020\** | *-0.14* | *[-0.25, -0.02]* |
| *fixed effect per PM task type* | *Focal* | *35.07* | *<.001\*\*\** | *1.68* | *[1.58, 1.77]* |
| | *Non-Focal* | *-1.10* | *.270* | *-0.08* | *[-0.21, 0.06]* |
| | *Time-based* | *-2.21* | *.030\** | *-0.16* | *[-0.29, -0.02]* |
| mixed effect per Interaction between Delay and PM task type | Early Focal | 3.03 | <.001\*\*\* | 0.31 | [ 0.11, 0.51] |
| | Late Focal | 2.56 | .010\*\* | 0.27 | [ 0.06, 0.47] |
| | Early Non-Focal | 2.66 | .010\*\* | 0.28 | [ 0.07, 0.48] |
| | Late Non-Focal | 1.38 | .170 | 0.15 | [-0.06, 0.36] |
| | Early Time-based | 2.42 | .020\* | 0.25 | [ 0.05, 0.46] |
| | Late Time-based | 17.75 | <.001\*\*\* | 1.39 | [ 1.23, 1.54] |

The effects of the best model are displayed in *Italics*; CI = Confidence interval.

The post hoc comparisons of the least-squares means of the models' predictors indicated that there is not a significant difference between the types of PM tasks. In contrast, a significant difference in the z ratio = 2.38, p = .018 was detected between the two delay lengths, suggesting that the effect of each delay on PM functioning is significantly different from each other. This finding hence confirms that the shorter delay tends to improve PM functioning, while the longer delay tends to worsen PM functioning. A similar pattern was also revealed for the significant difference of the z ratio = 3.03, p = .029 between the least-squares means of the early focal task and the late time-based task. Hence, an early focal task affects everyday PM functioning positively, and it is significantly different from the negative effect of a late time-based task on everyday PM functioning.

*Everyday prospective memory ability*

The models for everyday PM ability are presented in Table 5. Compared to the null model, the models only containing the length of delay, the PM task type, or the interaction between them as predictors were found to explain everyday PM ability significantly better. The delay model appeared substantially better than the PM task type model. Nevertheless, the interaction model was found to be the best model with a substantially better AIC, deviance, and $R^2$. Notably, the interaction model was found to explain 27% of everyday PM ability.

Table 5. Mixed effects generalised linear regression models for the dichotomous PM score.

| Model | F | $\chi^2$ | p-value ($\chi^2$) | AIC | Deviance | $R^2$ |
|---|---|---|---|---|---|---|
| Null (only random effects) | - | - | - | 344.44 | 340.44 | 0.00 |
| Delay | 24.48 | 25.34 | <.001*** | 321.11 | 315.11 | 0.13 |
| PM task type | 07.01 | 14.64 | <.001*** | 333.80 | 325.80 | 0.08 |
| *Interaction between Delay and PM task type* | *07.67* | *47.39* | *<.001*** | *307.06* | *293.06* | *0.27* |

The best model is displayed in *Italics*; AIC = Akaike's Information Criterion; For AIC and Deviance, a smaller value indicates a better fit for the model; The displayed $\chi^2$ statistics are against the null model; $R^2$ is a pseudo-$R^2$ for mixed effects generalised linear regression models.

The effects on everyday PM ability are presented in Table 6. In contrast to everyday PM functioning, the random effects were not found to significantly affect everyday PM ability. However, similarly to PM functioning, a greater length of delay was found to decrease everyday PM ability scores, while the shorter delays appear to positively affect everyday PM ability. Likewise, the focal event-based tasks facilitate improved PM ability, while the non-focal and time-based tasks appear to decrease everyday PM ability scores. This finding further supports the importance of the focality and salience of the target cue in everyday PM ability. Lastly, all the combinations between the type of task and the length of delay were found to have a significant effect on everyday PM ability.

Table 6. Models' effects on the dichotomous PM Score.

| Effect | Levels | z Value | p-value (>|z|) | β coefficient | β 95% CI |
|---|---|---|---|---|---|
| *Random effect per participant* | - | *0.76* | *0.44* | *0.10* | *[-0.15, 0.35]* |
| fixed effect per Delay | Early | 3.86 | <.001*** | 0.78 | [ 0.38, 1.17] |
| | Late | -4.81 | <.001*** | -1.34 | [-1.88, -0.79] |
| fixed effect per PM task type | Focal | 3.20 | <.001*** | 0.77 | [ 0.30, 1.24] |
| | Non-Focal | -2.36 | .020** | -0.77 | [-1.41, -0.13] |
| | Time-based | -3.67 | <.001*** | -1.22 | [-1.87, -0.57] |
| *mixed effect per Interaction between Delay and PM task type* | *Early Focal* | *5.02* | *<.001*** | *3.00* | *[ 1.83, 4.17]* |
| | *Late Focal* | *4.05* | *<.001*** | *2.29* | *[ 1.18, 3.39]* |
| | *Early Non-Focal* | *4.38* | *<.001*** | *2.51* | *[ 1.38, 3.63]* |
| | *Late Non-Focal* | *2.03* | *.040** | *1.14* | *[ 0.04, 2.24]* |
| | *Early Time-based* | *4.22* | *<.001*** | *2.39* | *[ 1.28, 3.51]* |
| | *Late Time-based* | *-3.97* | *<.001*** | *-1.82* | *[-2.72, -0.92]* |

The effects of the best model are displayed in *Italics;* CI = Confidence interval.

Post hoc comparisons of the least-squares means of the models' predictors postulated that, between the types of PM tasks, only the focal event-based task has a significantly different effect on everyday PM ability from the time-based task, z ratio = 3.67, p <.001, while the remaining comparisons indicated no significant differences between the types of PM tasks. Also, in the same way as everyday PM functioning, a significant difference of a z ratio = 4.81, p <.001 was identified between the two lengths of delay, postulating that the improvement in PM ability due to shorter delays is substantially different from the decrease in everyday PM ability due to longer delays. Similarly, the early focal PM task had a significant different effect on everyday PM ability from the late non-focal PM task, z ratio = 3.65, p = .004, and the time-based task, z ratio = 5.02, p <.001. Also, the effects of the early non-focal task, z ratio = 4.38, p <.001, and the early time-based task, z ratio = 4.22, p <.001, are found to be significantly different from the effect of late time-based task on everyday PM ability. Finally, the effect of the late focal task was substantially different from the effect of the late time-based task, z ratio = 4.05, p <.001, on everyday PM ability. These findings postulate that the type of task does not appear as central as the length of delay, which appears to affect more the non-focal and time-based tasks with less focal and salient PM cues. However, the effect of delay is greater in time-based tasks compared to non-focal tasks. This suggests that the different effect of delay on non-focal and time-based tasks is due to the focality of the PM cue.

*Pairwise multiple comparisons of mean rank sums*

Box-violin plots were preferred to visualise the comparisons because they also illustrate the kernel probability density of the data. This allowed an inspection for ceiling effects, which there did not appear to be for PM performance. Figure 3 demonstrates that there was a marginally significant difference between the performance on focal and non-focal event-based tasks for everyday PM functioning, while the difference between non-focal and time-based tasks was not significant. On the other hand, significantly decreased performance was found on time-based tasks compared to focal event-based tasks. Figure 4 illustrates that the significant differences and the absence of significant differences for everyday PM ability are comparable to the outcomes for everyday PM functioning. The only difference for PM ability is that the performance on non-focal tasks appears to be significantly, instead of marginally, different compared to performance on focal tasks.

Figure 3. Prospective memory functioning per type of task

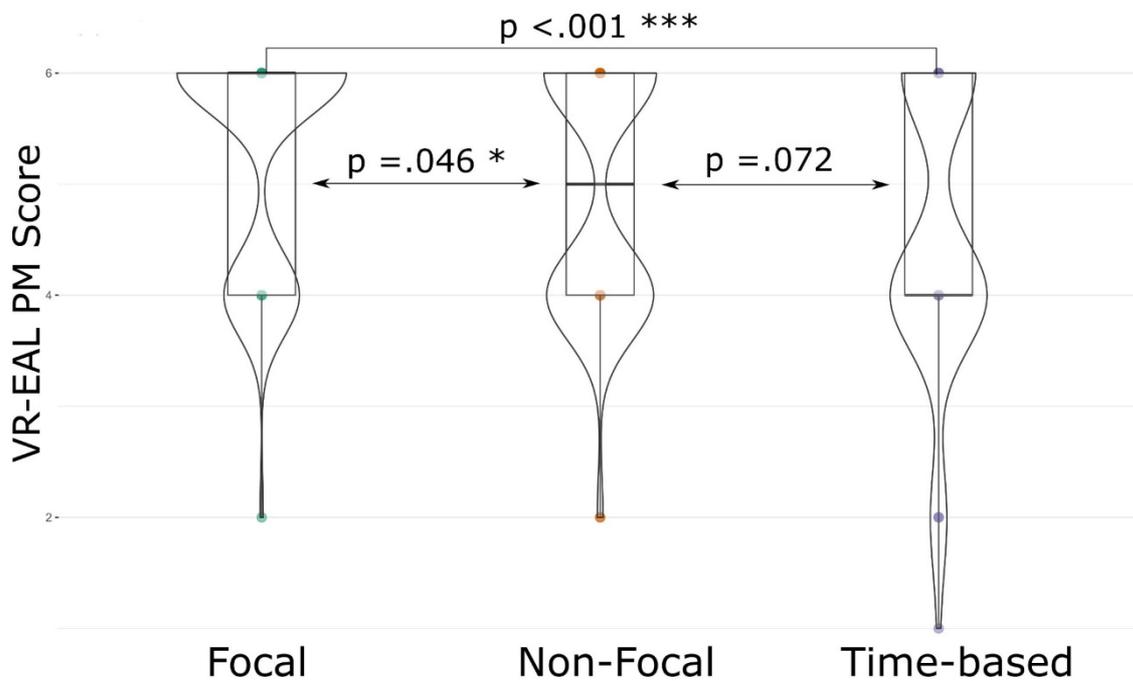

Figure 4. Prospective memory ability per type of task

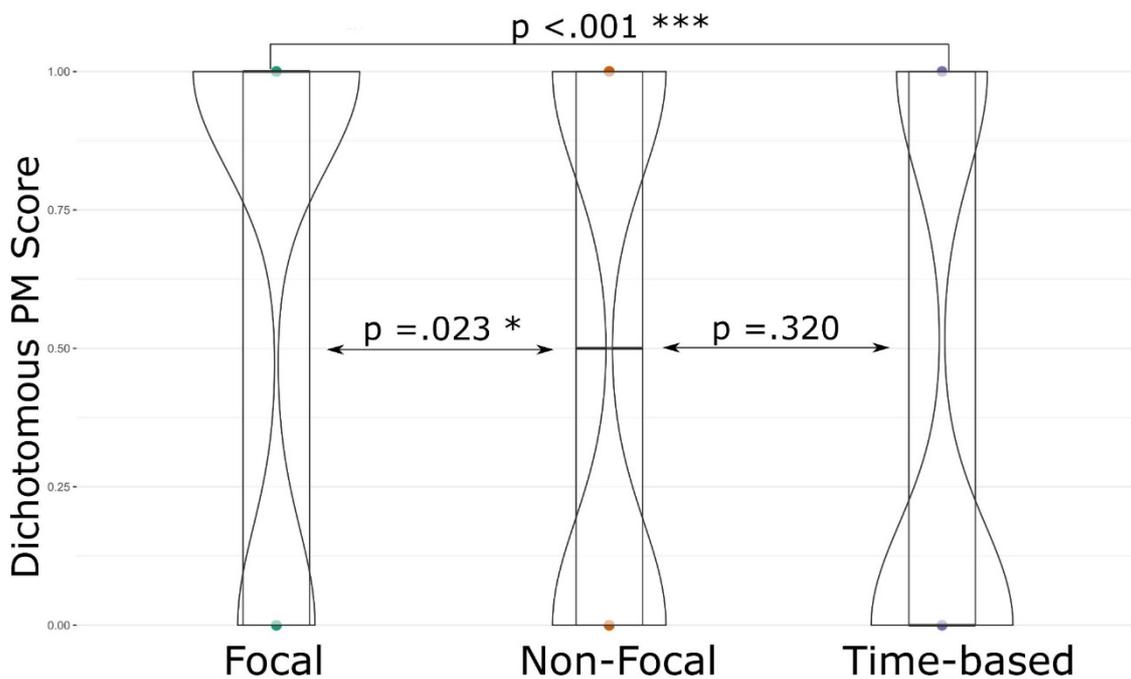

Regarding the effect of delay on both everyday PM functioning (see Figure 5) and ability (see Figure 6), there was decreased performance on the PM tasks performed at the late stage of the VR-EAL scenario compared to the tasks performed at an early stage of the scenario. Also, for both everyday PM functioning (see Figure 7) and ability (see Figure 8), there was not any significant difference in the performance between focal, non-focal, and time-based tasks which were performed at an early stage of the VR-EAL scenario. On the contrary, there were noticeable differences between the PM tasks when they were performed at a late stage of the scenario. The comparisons, for both everyday PM functioning and ability, showed that the performance on late focal tasks is substantially greater than the performance on late non-focal tasks, and performance on both late focal and non-focal event-based tasks is significantly greater than performance on late time-based tasks. Finally, the diverse effect of the length of delay on each type of task was also apparent in the comparisons between the tasks of the same type, which were performed at an early and late stage of the scenario. For both everyday PM functioning (see Figure 9) and ability (see Figure 10), there was a significant difference on the performance between early non-focal and late non-focal, as well as between early time-based and late time-based, where the performance on the late tasks was substantially poorer. However, for both everyday PM functioning and ability, there was not any significant difference in the performance between early focal and late focal tasks.

Figure 5. Prospective memory functioning per delay

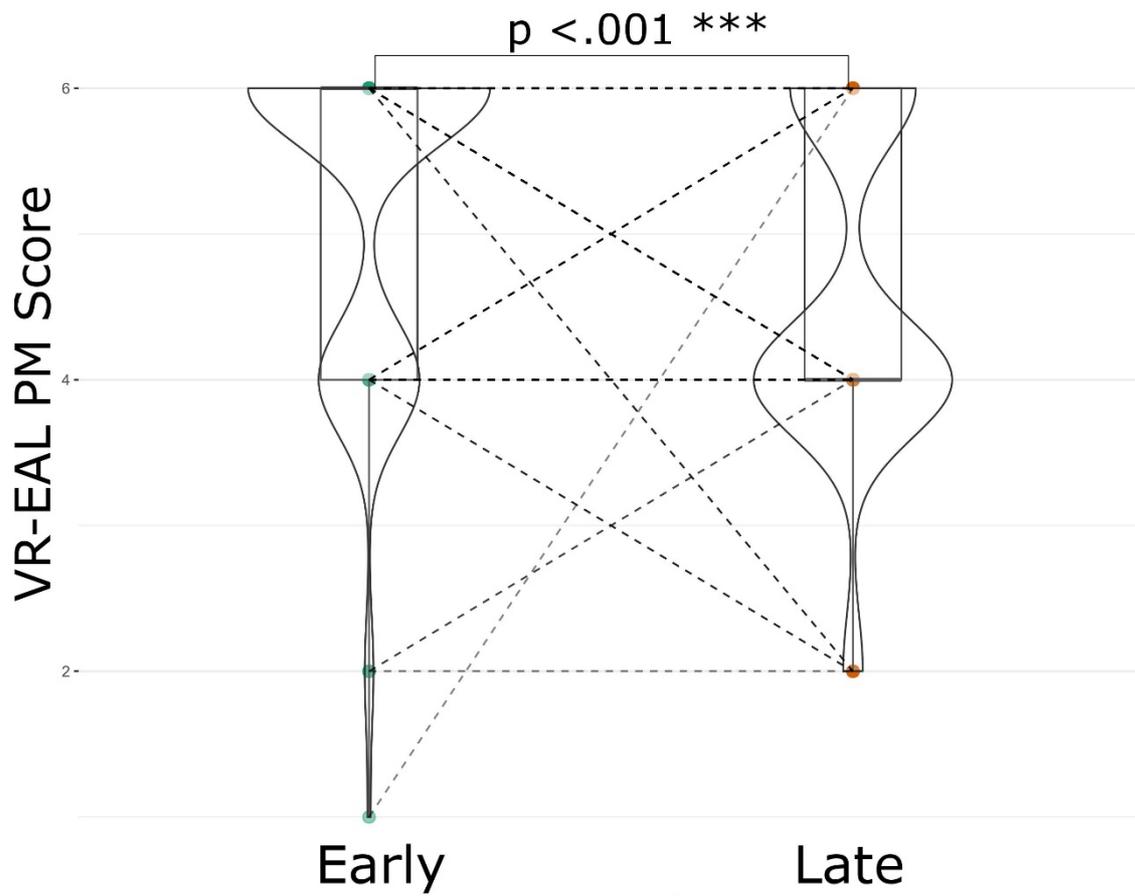

Figure 6. Prospective memory ability per delay

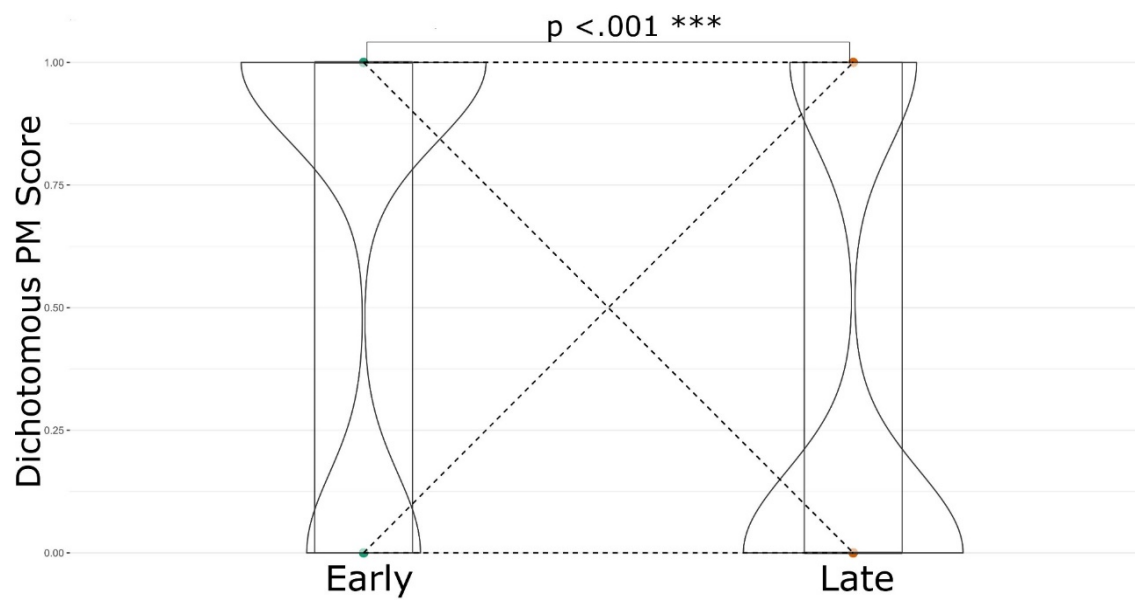

Figure 7. Prospective memory functioning per task type and delay

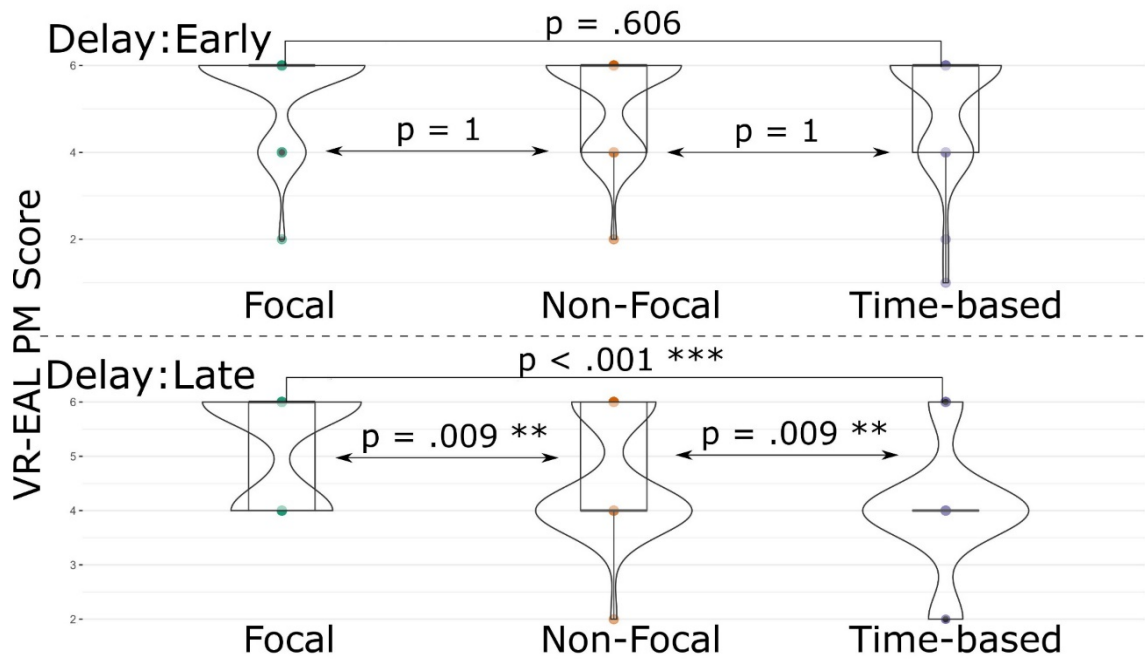

Figure 8. Prospective memory ability per task type and delay

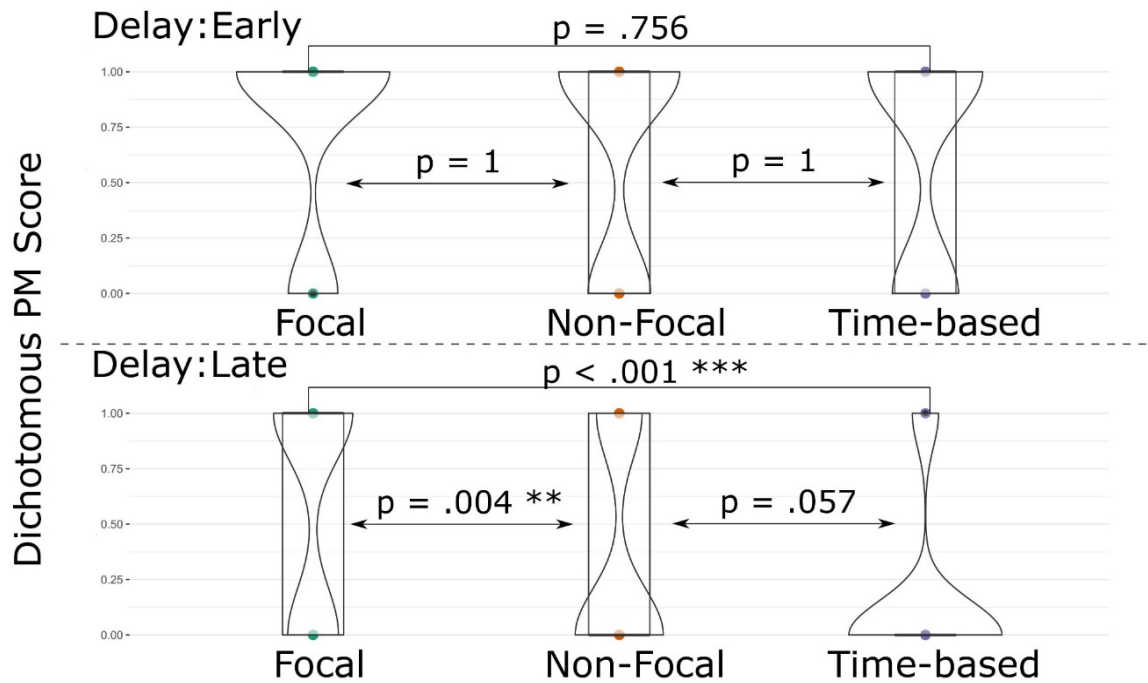

Figure 9. Prospective memory functioning per delay and task type

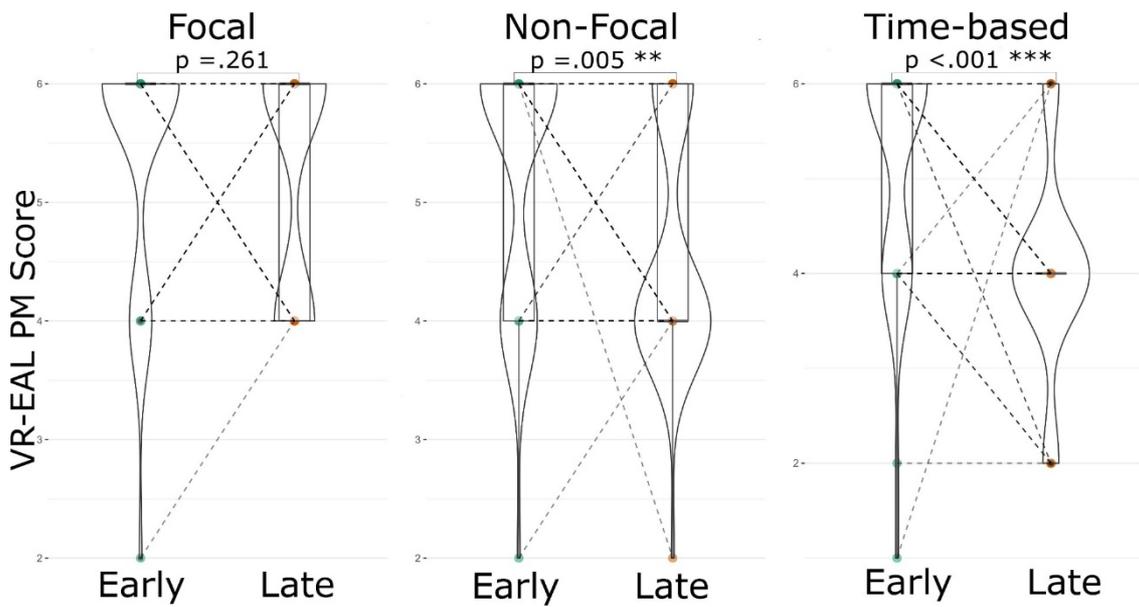

Figure 10. Prospective memory ability per delay and task type

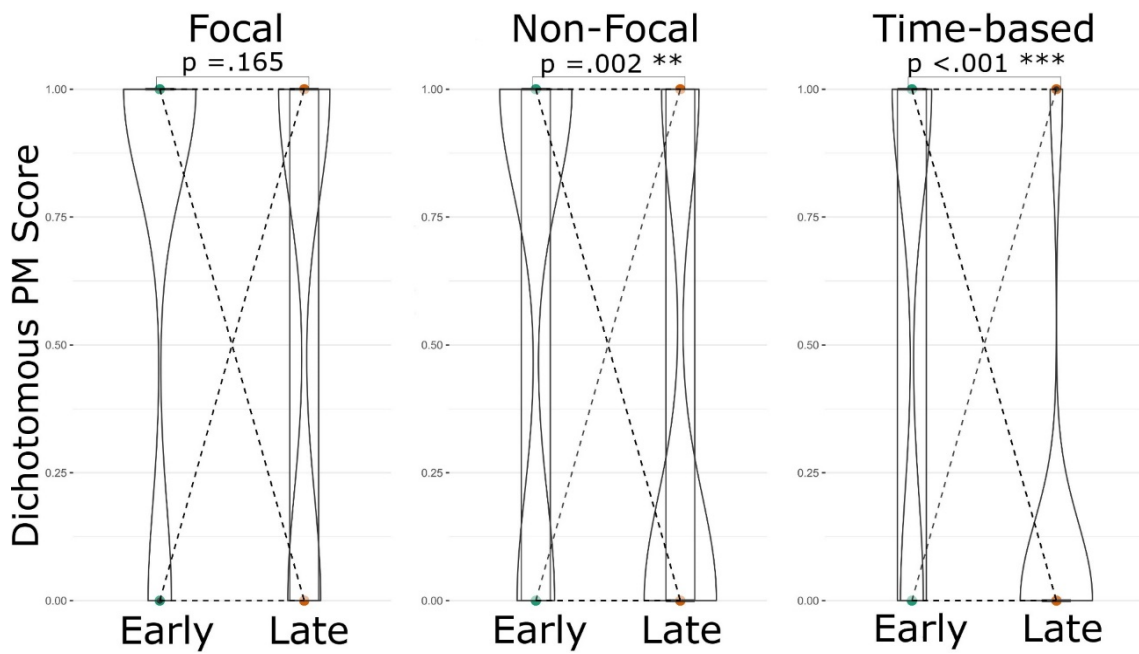

**Discussion**

This study aimed to investigate the effects of delay and PM task type on everyday PM performance by using an immersive VR paradigm with enhanced ecological validity. The inability of laboratory experiments to generalise their findings to everyday PM performance has caused several discrepancies with the observations derived from naturalistic approaches. The findings of the current study further postulated the unreliability of the findings of laboratory tasks for depicting everyday PM performance. Overall, the length of delay was found to have a greater effect than the type of PM task type on everyday PM functioning and ability. However, the effect of the length of the delay significantly differed in terms of the performance on focal, non-focal, and time-based tasks.

We differentiated everyday PM performance in terms of everyday PM functioning (i.e., performing everyday PM tasks using external aids) and everyday PM ability (i.e., performing everyday PM tasks without using external aids). This differentiation may be useful to appraise the added value of using external aids, and judge whether an individual is capable of performing everyday PM tasks either independently, or by using compensatory strategies. In addition, an impairment in PM functioning may indicate that an individual requires support (e.g., by a support worker or a care assistant) to perform everyday PM tasks.

*Everyday prospective memory performance*

The length of the delay was found to explain everyday PM functioning better than the type of PM task, although both were found to be predictors in the best model. Nevertheless, the length of delay together with the type of PM task type only explained 5% of the observations. The current findings contradict the multiprocess framework which states that the type of PM task (e.g., focal or non-focal) is the significant contributor to PM performance (McDaniel & Einstein, 2007). However, the claims of the multiprocess framework are based on the findings of non-ecologically valid laboratory tasks (e.g., Conte & McBride, 2018; Einstein *et al.*, 1997, 2000; McBride & Flaherty, 2020; Mullet *et al.*, 2013; Scullin *et al.*, 2010; Smith, 2003; Smith *et al.*, 2007; Zuber *et al.*, 2019). As mentioned earlier, discrepancies between naturalistic and laboratory tasks have previously been found, including the age-prospective memory-paradox, and postulated the necessity of using ecologically valid tasks to explain everyday PM performance (Kvavilashvili *et al.*, 2013; Niedźwieńska & Barzykowski, 2012; Phillips *et al.,* 2008; Uttl, 2008).

Nonetheless, these discrepancies may be attributed to individual differences and the use of compensatory strategies and external memory aids, which are common in daily life for performing PM tasks (Phillips *et al.,* 2008; Shelton & Scullin, 2017; Uttl, 2008). In our study, random effects were found to significantly affect everyday PM functioning (i.e., also using external aids such as prompts), postulating the influence of individual differences on everyday PM functioning. These findings seem better aligned with the dynamic multiprocess framework, where the contextual variability (similar to everyday life) and individual differences are considered as factors which may affect how individuals perform PM tasks (Scullin *et al.*, 2013; Shelton & Scullin, 2017).

Nevertheless, in contrast with everyday PM functioning, individual differences (i.e., random effects per participant) did not appear to significantly affect everyday PM ability (i.e., performing the tasks without using external aids such as prompts). This postulates that individual differences may predominantly pertain to the use of compensatory strategies and external aids (see also Phillips *et al.,* 2008; Uttl, 2008). Moreover, compared to the effect of PM task type, the length of delay again appeared to better explain everyday PM performance. However, in the case of everyday PM ability, the effect difference between the length of delay and type of PM task was smaller compared to the effect difference observed in everyday PM functioning. Nevertheless, these findings again contradict the findings commonly found in laboratory designs, where the effect of the PM task type is central. In our study, however, the interaction between the length of delay and the PM task type was found to better explain everyday PM ability. Notably, this interaction explained 27% of the observations pertinent to everyday PM ability.

*Effect differences by the length of delay*

The post-hoc comparisons of the least-squares means indicated that the increase in everyday PM functioning and ability due to the short delays of 15-30 minutes is substantially different from the decrease in everyday PM functioning and ability due to the long delays of more than 45 minutes. This significant difference further suggests the important role of the length of delay on everyday PM performance.

On the contrary, there were no significant differences between the effects of focal, non-focal, and time-based tasks on everyday PM functioning. This contradicts the findings of laboratory experiments, where the focal event-based tasks (i.e., positive effects) had significantly different effects on PM performance than the non-focal event-based tasks (i.e., negative effects; e.g., McDaniel, Shelton, Breneiser, Moynan, & Balota, 2011; Mullet *et al.*, 2013; Scullin *et al.*, 2010; Zuber *et al.*, 2019), and the effects of both event-based types were substantially different from the effects of time-based tasks (i.e., negative effects; e.g., Conte & McBride, 2018; Zuber *et al.*, 2019). As these discrepancies were also seen in naturalistic experiments, they may be due to the effect of using external aids (e.g., notes and prompts), which assist PM performance in ecologically valid paradigms (Phillips *et al.,* 2008; Uttl, 2008). These external aids remind the individual of the intended action (Phillips *et al.,* 2008; Uttl, 2008). However, there are individual differences on how well external aids are used (Phillips *et al.,* 2008; Shelton & Scullin, 2017; Uttl, 2008). Hence, performance with external aids may indicate examinee's PM functioning, as is observed in naturalistic tasks, but not examinee's PM ability, which is observed in lab-based tasks.

Nevertheless, only one significant difference between the effects of diverse PM task types on everyday PM ability was found. The effect of focal event-based tasks (i.e., increasing performance) on everyday PM ability was found to be substantially different from the effect of time-based tasks (i.e., decreasing performance) on everyday PM ability. The absence of significantly different effects of diverse PM types (i.e., focal vs non-focal, non-focal vs time-based) on everyday PM ability (i.e., without using prompts) is again discrepant to the findings of laboratory experiments. However, the effect difference between time-based vs focal event-based observed in our study is in line with the corresponding findings of other laboratory experiments (e.g., Conte & McBride, 2018; McBride & Flaherty, 2020; Zuber *et al.*, 2019). Nevertheless, overall, the findings pertaining to the differences between the effects of diverse PM task types on everyday PM performance do not align with the findings of laboratory experiments. Since the discrepancies were found for both everyday PM functioning and ability, this suggests that these differences are due to the enhanced ecological validity of the assessment used in the current study.

However, as mentioned above, the length of delay had a significant effect on everyday PM performance. Considering the different focality of the PM cue for each PM task type, the length of delay may have differently affected PM performance on each type of task. The post-hoc comparisons of the least-squares means revealed a significant difference between the effects of the early focal event-based task (i.e., positive) and the late time-based task (i.e., negative) on everyday PM functioning. In contrast with the absence of significant differences between the effects of diverse PM task types, this finding further postulates the importance of delay on PM performance. However, the use of prompts may have mitigated the effect of delay on the performance on each PM task type.

Indeed, several significant differences among the effects of distinct interactions between delay and PM task types on everyday PM ability were observed. The early focal PM task had a significant different effect (i.e., positive) on everyday PM ability from the late non-focal PM task, and the time-based task (i.e., a negative effect). In addition, the positive effects of the early non-focal task and the early time-based task are found to be significantly different from the negative effect of late time-based task on everyday PM ability. Since the increasing or decreasing nature of the effect was defined by the length of delay, these findings further support the important role of delay length. However, the positive effect of the late focal task was substantially different from the negative effect of the late time-based task on everyday PM ability. This finding postulates that PM performance on focal event-based tasks is affected significantly less than PM performance on time-based tasks with a long delay of more than 45 minutes.

*Performance differences*

The findings from the performance comparisons also supported the importance of the length of delay. As with the post-hoc comparisons of the least-squares means, there was a significant difference between the tasks performed at an early (i.e., 15 – 30 mins) and late (i.e., 45 – 60 mins) delay within the scenario for both everyday PM functioning and ability. However, the performance comparisons indicated some significant differences, which were not observed via the post-hoc least-squares means comparisons. For both everyday PM functioning and ability, the performance on focal event-based tasks was significantly better than the performance on non-focal event-based tasks. Yet, the significance of this performance difference was marginal. However, the performance on time-based tasks was substantially worse than the performance on focal event-based tasks, while the performance on non-focal and time-based tasks did not significantly differ. Nevertheless, our work suggests that one should also consider the effect of delay in these comparisons to allow one to draw reliable conclusions.

Indeed, for both everyday PM functioning and ability, there were no significant performance differences between the types of PM tasks when the length of delay was 15 – 30 minutes. These findings speak against the findings of laboratory-based experiments (e.g., Conte & McBride, 2018; McBride & Flaherty, 2020; McDaniel *et al.*, 2011; Mullet *et al.*, 2013; Scullin *et al.*, 2010; Zuber *et al.*, 2019), but support the findings of Niedźwieńska and Barzykowski, (2012) in a naturalistic setting. Importantly, the laboratory PM tasks are commonly performed after a delay length of 10–15 minutes, which is a considerably shorter delay compared to the PM tasks performed during the early stage of the VR-EAL scenario (i.e., after 15 – 30 minutes).

Nevertheless, for both everyday PM functioning and ability, there were significant performance differences when the length of delay was 45 minutes or more, which differs from Niedźwieńska and Barzykowski (2012). Niedźwieńska and Barzykowski (2012) did not report how their participants performed the naturalistic PM tasks, and there were no on-going tasks except to watch the news on television. Furthermore, scoring was dichotomous (i.e., 0 = failure, 1 = success), which may have resulted in ceiling effects, and they did not consider the use of external aids (e.g., a prompt by their spouse to call the experimenter). Lastly, the on-going task favoured the non-focal task because the target cue (i.e., the name of a politician) was typically mentioned during the first couple minutes of the news, while the focal cue (i.e., the weather map of Poland) was usually at the end of the news. Some of these differences may explain the discrepancy between our findings and those of Niedźwieńska and Barzykowski (2012).

In our study, when the length of delay was more than 45 minutes, performance on the focal event-based task was significantly better than performance on the non-focal event-based and the time-based task for both everyday PM functioning and ability. In addition, the performance on non-focal task was better than the performance on time-based task for both everyday PM functioning and ability. The performance difference between them was significant for PM functioning, while it was marginal (i.e., $p = .057$) for PM ability. Given the lack of differences among the PM tasks during the early stage of the VR-EAL scenario, the differences at the late stage of the scenario suggest that the length of delay differently affects performance on each type of task.

The salience of the PM cue does not appear adequate to explain these performance differences across PM tasks, since cue salience is comparable across non-focal and time-based tasks, especially in the case of the VR-EAL. Instead, the focality of the PM cue might be the reason for these differences, since the target cue's focality is lower in non-focal event-based and time-based tasks, with time-based tasks (e.g., at 1 pm – take your meds) having even lower focality than non-focal event-based tasks (e.g., when you return – give an extra pair of keys to Alex). Another explanation for the performance differences between non-focal and time-based tasks after a late delay may be the time monitoring that is required for the time-based task. Individuals do not implement constant time monitoring, which is cognitively demanding and tiring (Cona *et al.,* 2012; Kliegel *et al.,* 2001; Mäntylä *et al.*, 2007). The individuals predominantly adopt a strategic time monitoring approach, where after checking the time, they estimate when they should check the time again and regulate their frequency of time monitoring (Cona *et al.,* 2012; Kliegel *et al.,* 2001; Mäntylä *et al.*, 2007). Strategic monitoring appears to be more efficient and less cognitively demanding, while it does not seem to be affected by a delay (Cona *et al.,* 2012; Kliegel *et al.,* 2001; Mäntylä *et al.*, 2007). Therefore, the focality of the cue appears to be a more reasonable explanation for the significant effects of delay on everyday PM performance.

   Moreover, the performance on all types of late PM tasks was poorer than the performance on all types of early PM tasks. Our findings are concordant with the findings of previous studies that have attempted to examine the effect of delay on focal (e.g., Kliegel & Jager, 2006; Meier *et al.*, 2006; Scullin & McDaniel, 2010) and non-focal event-based tasks (e.g., Martin *et al.*, 2011; McBride, Beckner, & Abney, 2011), as well as time-based tasks (McBride, Coane, Drwal, & LaRose, 2013). It has been argued that a delay does not affect PM performance or even increases performance if the ongoing task permits rehearsal (Hicks, Marsh, & Russell, 2000; Kliegel & Jager, 2006; Martin *et al.*, 2011). However, the results of our study revealed a substantial decline in PM performance due to the delay. In VR-EAL, the ongoing tasks attempt to be highly engaging and demanding like real-life tasks (e.g., cooking, shopping, searching for items, and paying attention to the road), so they are not thought to permit rehearsal. This might explain why there was not a practice effect and that the duration of the delay negatively affected PM performance. Our findings are more in line with studies where the duration of the delay (e.g., 12 hours) is more representative of everyday PM (e.g., McBride *et al.*, 2013; Scullin & McDaniel, 2010). Also, the ongoing tasks are equally demanding at both the early (e.g., cooking and searching for items) and late (e.g., shopping and paying attention to the road) stages of the scenario. Hence, they seem equally disruptive of the early and late PM tasks.

   In VR-EAL, the focality of the PM cue (i.e., the semantic association with the intended action) was stronger in focal event-based tasks (e.g., a sign displaying the word "library" when the action was to return a book), weaker in non-focal tasks (e.g., having breakfast when the action was take the meds), and even weaker in time-based tasks (e.g., detecting that the time is 1 pm when the action is to take your meds). As discussed above, the focality of the PM cue seems to explain the differential effect of delay on each PM task type. These results show similarities with the encoding and retrieval of episodic memory, where stronger semantic associations between the items to be memorised, as well as the stronger associations between the cue (i.e., reminder) and the memorised items, assist with better encoding and retrieval of the episodic memory (Frankland, Josselyn, & Köhler, 2019; Tulving, 1983; Tulving & Thomson, 1973; Xue, 2018). Indeed, the pattern of PM forgetting seems comparable to episodic memory forgetting on non-focal event-based (Martin *et al.*, 2011; McBride *et al.*, 2011) and time-based tasks (McBride *et al.*, 2013), which further suggests that the encoding and retrieving mechanism is similar in PM and episodic memory and emphasizes the importance of cue focality (i.e., the semantic association between PM cue and intended action). However, the role of the PM's cue focality should be further investigated and differentiated from the role of cue salience.

Moreover, VR-EAL is a neuropsychological battery that attempts to simulate everyday life. The tasks are performed in a realistic order (e.g., first preparing the breakfast, and then driving to the supermarket) in an attempt to simulate real-life (Kourtesis et al., 2020b). While it is possible that the VR-EAL is susceptible to order effects, given that a selective decline in performance was only observed for non-focal event-based and time-based tasks, but not focal event-based tasks or the other everyday cognitive function tasks, it is unlikely that this is the case (see also Kourtesis et al., 2020a). Nevertheless, the effect of delay proportional to the cue attributes should be further explored.

### *Methodological considerations for future prospective memory studies*

#### *Differentiating functioning from ability*

In the current study, everyday PM functioning (i.e., using external aids) and ability (i.e., without requiring external aids) revealed similar performance differences on PM tasks. However, delay length appeared to have a greater negative effect on PM ability compared to PM functioning. Delay length may have had less of an effect on PM functioning because of the use of external aids (Phillips *et al.,* 2008; Shelton & Scullin, 2017; Uttl, 2008). One should consider functioning and ability when assessing PM, as functioning better describes individuals' PM performance in the real-world, and this distinction may have contributed to the discrepancies between the findings of naturalistic and laboratory experiments.

#### *Differentiating the cue's focality from the cue's salience*

In the current study, the effect of delay was attributed to the focality of the target cue for each type of PM task. However, it is difficult to consider salience and focality separately as the cue's features are typically coupled together (i.e., high focality with high salience or low focality with low salience) in PM research. This coupling of both features prevents one from examining their separate contributions on PM performance. For example, the salience of the target cue may affect the type of monitoring approach adopted (e.g., intentional or passive), while the focality of the cue may affect the encoding and retrieval of the PM intention. Future work might attempt to separate salience and focality in order to elucidate the individual roles that these cue features play on everyday PM performance.

#### *Ecological validity*

The discrepancy between the current results and the results of previous laboratory-based PM studies further highlights the importance of ecological validity for understanding everyday PM performance. The VR-EAL provides a testing environment that is closer to real life situations. Yet, while daily life simulations like VR-EAL are suitable for the investigation of real-life PM, they may also be susceptible to confounding factors and fail to examine specific processes. For example, an environmental stimulus (e.g., a building with several signs) or an interaction with a 3D character (i.e., requirement of communicational skills) may affect the PM performance, while they were not experimentally controlled for it. Laboratory tasks have the advantage of isolating confounding factors and allowing the examination of specific processes. However, laboratory PM tasks in their current form suffer from certain limitations (e.g., a two-dimensional environment, keyboard-based responses, static stimuli, and a lack of realism). In addition, the interface of laboratory PM tasks is keyboard-based. Hence, familiarity with computer systems might affect performance (Zaidi et al., 2018).

In contrast, gaming and computing competency does not affect performance in VR systems, when they have ergonomic and naturalistic interactions (Kourtesis et al., 2020a, 2020b; Zaidi et al., 2018). Moreover, older adults have shown high acceptance of immersive VR systems (Huygelier et al., 2019; Syed-Abdul et al., 2019). Hence, one solution would be to modernize existing laboratory tasks using immersive VR technology, which would enable the use of a 360º environment with dynamic and realistic stimuli, and naturalistic and ergonomic interactions within the testing environment. Hence, the implementation of immersive VR laboratory tasks may reduce the divergence from real-life conditions, and enable a thorough examination of PM performance, components, and cue attributes.

*Limitations and future studies*

This study has some limitations that should be considered. While our sample size allowed us to perform sound statistical analyses, future studies should include a larger sample for a more extensive investigation of the factors influencing PM performance (e.g., diverse cue attributes and interactions). Furthermore, the participants were only young adults. Future work should also include a more diverse population, including both younger and older adults, which will facilitate the examination of the effect of aging on everyday PM performance using our VR-EAL environment. Finally, the VR-EAL has the limitation that it cannot directly measure time monitoring so a future version of the VR-EAL should integrate such a mechanism.

*Conclusions*

Everyday PM performance, in terms of both functioning and ability, appears to be affected more by the length of the delay between the encoding and retrieval of the PM intention, rather than the type of PM task performed (i.e., focal event-based, non-focal event-based, and time-based). Nevertheless, the effect of the delay length differentially affected performance on focal, non-focal, and time-based tasks. This was proportional to the focality of the PM target cue (i.e., semantic relationship between the cue and the intended action). In conclusion, this study further highlighted the necessity of ecological validity to appropriately assess everyday PM functioning and ability. The implementation of immersive VR methods may facilitate an ecologically valid examination of everyday PM.